\documentclass[twocolumn]{revtex4}
\usepackage{hyperref}
\usepackage{amssymb,amsmath}
\usepackage{subfigure}
\usepackage{graphicx}
\usepackage{psfrag}
\newcommand{\be}[1]{\begin{equation}\label{#1}}
\newcommand{\ee}{\end{equation}}
\newcommand{\bea}{\begin{eqnarray}}
\newcommand{\eea}{\end{eqnarray}}
\newcommand{\eq}[1]{Eq.\,(\ref{#1})}
\begin{document}
\title{Polyatomic Molecules Formed with a Rydberg Atom in an Ultracold Environment}
\author{Ivan C. H. Liu}
\email{ivanliu@pks.mpg.de}
\author{Jan M. Rost}
\email{rost@pks.mpg.de}
\affiliation{Max-Planck Institute for the Physics of Complex Systems, N\"othnitzer 
Str.~38, 01187 Dresden, Germany}
\begin{abstract}
We investigate properties of ultralong-range polyatomic molecules formed
with a Rb Rydberg atom and several ground-state atoms whose distance from
the Rydberg atom is of the order of $n^2a_0$, where $n$ is the principle 
quantum number of the Rydberg electron. In particular, we put emphasis 
on the splitting of the energy levels, and elucidate the nature of the 
splitting via the construction of symmetry-adapted orbitals.
\end{abstract}
\maketitle
\section{Introduction}

The recent advancement of ultracold physics has made possible the study of
many interesting phenomena, ranging from the formation of molecules in a
Bose-Einstein Condensate (BEC) \cite{Wyn00,Ger00,Mcken02} to correlation
effects in ultracold neutral plasmas \cite{Kil99,Poh04}, where in the former
case the atoms are cooled to temperature in the nano-Kelvin range.  Combined
with narrow bandwidth lasers and high resolution spectroscopy \cite{Li03}, new
phenomena can be studied involving high-lying Rydberg states which have a
narrow spacing in energy of the order of 10 GHz.  One such example is the
theoretical prediction of the formation of ultralong-range dimers by a Rydberg
atom and a nearby ground-state atom \cite{Greene00}, the so-called
``trilobite'' molecules. The potential well supporting the vibrational bound
states is extremely weak compared to typical ground-state mole\-cules. The depth
of well is $\sim 15$GHz for $n$=30, where $n$ is the principle quantum number
of the Rydberg atom, and scales as $n^{-3}$. The long-range nature of such
system, bound at the equilibrium distance of the order of $10^3$a.u., is
rather unusual as well as the oscillatory feature at the bottom of the
potential well. This class of ultralong-range dimers should be distinguished 
from another kind, due to the Rydberg-Rydberg interaction \cite{Boi02}, 
whose molecular resonance was observed in the experiments \cite{Far03,Mar05}.

In this paper we address the question if more than one ground-state atom can
form, together with the Rydberg atom, a polyatomic molecule. We find, that
shape and symmetry of such polyatomic molecules follow a systematics which is
well understandable on the one hand side in terms of trilobite building
blocks, that is, linear combinations of wavefunctions which describe the
Rydberg atom and a single ground-state atom. From a more global perspective,
these polyatomic molecules can be classified according to irreducible
representations reflecting their symmetry properties, as it is well known in
quantum chemistry.

So far, trilobites have not been identified experimentally.  As the
experimentally achieved density of ultracold atomic ensembles increases
\cite{Seb05}, so does the likelyhood of detecting such molecular species.
However, there are still open questions concerning the realizability of such a
molecule under current experimental conditions. For example, the Rydberg
electron is very likely to interact with multiple ground-state perturbers, and
it is not so clear at first sight what the role of the Rydberg states is.
Such scenarios will also concern the proposed ``dipole blockade'' scheme for
quantum information processing \cite{Lukin01}. The characterization of
polyatomic molecules involving one Rydberg and several ground-state atoms
helps to understand the possible role of such ground-state perturber
better.

Technically speaking, we extend the calculations done by Greene, Dickinson and
Sadeghpour \cite{Greene00} to include multiple ground-state atoms using the
Fermi pseudo-potential treatment. More sophisticated methods exist
\cite{Chi02,Khu02}, but the results do not differ much, whilst the
qualitative features are certainly captured which suffice for the purpose of
the present article. Using the Fermi pseudo-potential also allows one to
calculate easily a large system.

We investigate, in particular, the effect of placing the perturbers in a
structured environment on the splitting of the adiabatic energy levels of the
molecular system. We will use group theory to obtain the total wavefunction of
the system in the framework of first-order perturbation theory via the
construction of the \emph{symmetry-adapted orbitals}.  For the sake of clarity
we restrict the investigation to atoms all lying in a plane (whereas the
Rydberg electron of course lives in the physical 3D space).  Atomic units are
used unless stated otherwise.

\section{The Hamiltonian}
First, we consider a ground-state atom with label $i$ located at distance
$R_i$ from the Rydberg atom. The ground-state atom influences the electron by
its polarization field, which has the form $-\alpha /2r^{4}$, where $\alpha$
is the atomic polarizability. For Rb atoms, the experimentally determined
value is $\alpha=319.2$ \cite{Molof74}. To a good approximation, the potential
-- extremely short-ranged with respect to the extension of a Rydberg
electronic wavefunction -- can be mimiced by the Fermi pseudo-potential,
namely~\cite{Fermi34},
\begin{equation}\label{fer}
\hat{V_i}=2\pi L[k_i]\delta (\vec{r}-\vec{R}_i).
\end{equation}
where $L[k_i]\equiv -(\mathrm{tan}\,\delta_{s})/k_i$ is the s-wave 
\emph{energy-dependent} scattering
length of the Rydberg electron colliding with a neutral atom;
$k_i^{2}/2=-1/2n^{2}+1/R_i$ its kinetic energy, and $\vec{r}$ its distance
from the mother ion. The s-wave phase shift
$\delta_s$ can be calculated using the modified effective range theory by
O'Malley \textit{et al}
\cite{Omalley61} and the zero-energy scattering length for triplet s-wave
calculated by Bahrim \textit{et al} \cite{Bahrim01}. The singlet scattering
length is much smaller, and hence we do not expect it to influence the
phenomena discussed here.

In the case of Rb atoms, the quantum defect is negligible for high $l$-states
($l\geqslant 3$), which are quasi-denergate. They are therefore well
represented by hydrogenic wavefunctions. The low-$l$ states, on the other
hand, split away from the $n$-manifold and do not interact with the high-$l$
states provided that the energy-dependent scattering length is sufficiently
small.  The high-$l$ states are also more interesting, because the Hilbert
space is larger in this case, which produces a more flexible system, i.e. the
different eigenstates are allowed to interfere with each other. This results
in a highly-polarizable complex. In this paper, we consider only the high-$l$
class.

The total Hamiltonian of the Rydberg electron interacting with its mother ion and 
$N$ ground-state atoms can be written as
\begin{equation}\label{Ham}
\hat{H}=\hat{H}_{0} + \hat{V}_N,
\end{equation}
where the ionic Hamiltonian is
\begin{equation}\label{Ham2}
\hat{H}_{0}=\frac{ \hat{p}^{2} }{2}-\frac{1}{r}.
\end{equation}
The potential $V_N$ is the $N$-fold sum over the interaction \eq{fer} with 
all ground-state atoms, i.e.,
\begin{equation}\label{manygs}
{\hat{V}}_{N} = 2\pi \sum_{i=1}^{N}L[k_i]\delta (\vec{r}-\vec{R}_i),
\end{equation}
where $i$ labels the $i$-th ground-state atom, and $N$ is the total number of
ground-state atoms.

The effect of $p$-wave electron scattering was neglected in our calculation,
however, in the dimer case it was studied by Hamilton {\it et al} \cite{Ham02}
and Khuskivadze {\it et al} \cite{Khu02}.

\begin{figure}
\unitlength=1cm
\includegraphics[width=1.0\columnwidth]{figures/E-R_2.eps}
\put(-6,5){\includegraphics[width=0.4\columnwidth]{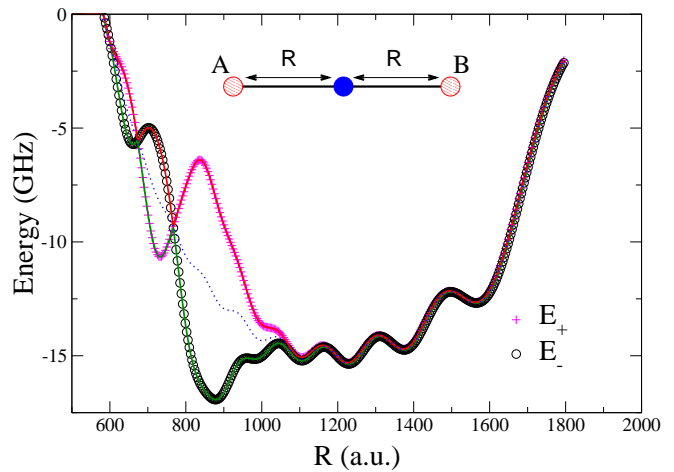}}
\caption{\label{fig:2gsbo}Adiabatic energy curves for the linear diatomic 
(dotted line) and triatomic molecule (solid lines) as a function of distance
$R$ between Rydberg and ground-state atoms, see also the sketch for the
triatomic molecule with the Rydberg atom in the middle. The solid lines are
the results from diagonalization, and the points ($+$) and ($\circ$) are the
enegry expectation values calculated from the symmetry-adapted orbitals
Eq.~\eqref{eq:+} and \eqref{eq:-}, respectively. The $E=0$ lies at the energy
of the $n=30$ manifold.}
\end{figure}

\section{Determination of Adiabatic Energy Curves}

Under the adiabatic approximation, the energy levels calculated from the
Hamiltonian $\hat{H}$ (including the perturbation) give, automatically, the
electronic structure of the molecular system involving $N+1$ atoms.  We
determine a cut through the Born Oppenheimer (BO) potential surface for
systems with $N=2$, 3, and 4 ground state atoms which are uniformly placed on
a circle with radius $R$ centered at the Rydberg core. These three cases
correspond to linear, triangular, and square geometries, respectively. The cut
we choose corresponds to the breathing mode, i.e. $R$ is varied. We calculate
the BO curves using two methods: (i) the direct diagonalisation of $\hat{H}$;
and (ii) the \emph{projection operator method} \cite{Levine75} to construct
symmetry-adapted orbitals and determine the BO curves from standard
perturbation theory. Both methods are accurate, but the latter gives a deeper
insight into the quantum mechanical properties such as the energy degeneracy.

The Fermi pseudo-potential is usually valid for $n\gtrsim$25-30. For smaller
principle quantum numbers, the scattering of $e^{-}+\textrm{Rb}$ and the
polarization of the neutral perturber by the Rydberg atomic core are not
independent of each other \cite{Fab86}.  Here, we present calculations for
$n=30$, which also allows us to compare our results directly with that
previously obtained for the Rb$_2$ dimer \cite{Greene00,Khu02}.

The eigenvalues from the unperturbed Hamiltonian, $\hat{H}_0$,
yields simply the hydrogenic energy $E_0=-1/2n^{2}$, so for convenience we
set this to be zero throughout this article.

\section{\label{sc:2gs}Collinear Triatomic Molecule (N=2)}

Consider two ground-state atoms (A and B) placed on either side of a Rydberg
atom with distances $R$ forming a collinear triatomic molecule.
This configuration corresponds to $N$=2, and the numerical result of the
BO curves are plotted in Fig.~\ref{fig:2gsbo}. 

In order to understand the splitting of the energy levels, we use the
perturbed state $|\psi_{n}(\vec{r})\rangle$ when only one of the two
ground-state atoms is present as the building block for constructing the total
electronic wavefunction. The perturbed state can be can be explicitly written
as \cite{Omont77}
\begin{equation}\label{tri}
|\psi_{n}(\vec{r})\rangle=\sum_{q}{ \phi^{*}_{nq}(\vec{R}) |\phi_{nq}(\vec{r})\rangle},
\end{equation}
where the index $q$ runs over all the degenerate states which includes all
$l$'s and $m$'s with $l\geq3$. We call this wavefunction the ``trilobite''
wavefunction since it produces the probability density like that drawn in
Ref.~\cite{Greene00}. The two wavefunctions, which clearly satisfy the
parity of the collinear geometry, are
\begin{equation}
|\psi_{+}(\vec{r})\rangle=|\psi^{A}_{n}(\vec{r})\rangle+
|\psi^{B}_{n}(\vec{r})\rangle\label{eq:+},
\end{equation}
and
\begin{equation}
|\psi_{-}(\vec{r})\rangle=|\psi^{A}_{n}(\vec{r})\rangle-
|\psi^{B}_{n}(\vec{r})\rangle\label{eq:-},
\end{equation}
where the superscript A and B are the labels of the ground-state atoms; 
$|\psi^{A}_{n}(\vec{r})\rangle$ and $|\psi^{B}_{n}(\vec{r})\rangle$ are the
trilobite wavefunctions when only atom A or B is present. By choosing the projection
axis $\hat{z}$ so that it aligns with the internuclear axis, the only degenerate
states that contribute are those with non-zero value along $\hat{z}$.
They are in this case the states with $m=0$.

Using the above ansatz to calculate the expectation value
$\langle\hat{V}_N\rangle$ yields immediately two energies $E_+$ and $E_-$,
which are distinguished by their parities:

\begin{align}
E_{\pm}(R) = L[k]\sum^{n-1}_{l=3}{(2l+1)|u_{nl}(R)|^{2}}\left\{
\begin{array}{l}
\text{$l=$ even for $E_+$},\\
\text{$l=$ odd for $E_-$},
\end{array}\right.
\end{align}
where $L[k]=L[k_A]=L[k_B]$, $R=|\vec{R}_A|=|\vec{R}_B|$ and $u_{nl}$ is the radial part of
the hydrogenic wavefunction $\phi_{nl}(\vec{R})=u_{nl}(R)Y_{lm}(\theta,\varphi)$.

\begin{figure}
\includegraphics[width=1.0\columnwidth]{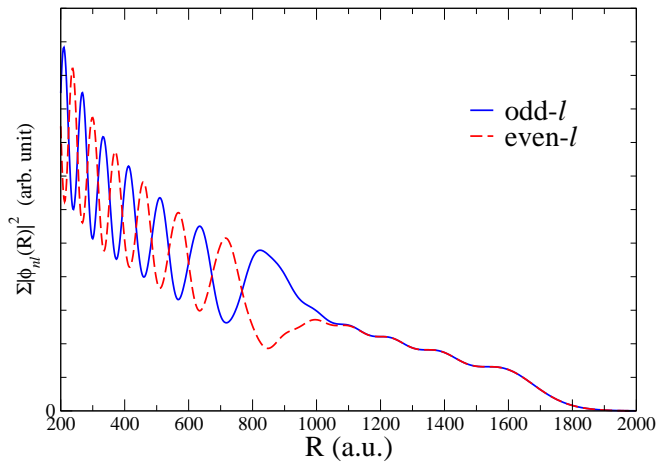}
\caption{\label{fig:sumpd}Sum of probability densities
$\sum_l|\phi_{nl}(\vec{R})|^2$ for states with even and odd angular momentum
components $l$ as a function of the radial distance.}
\end{figure}

Hence, we see that, with the inclusion of the second perturber, two curves
split away from the $n$-manifold, with one corresponding to \emph{gerade} and
the other to \emph{ungerade} symmetry. They both converge at large distance to
the curve when only one ground-state atom is present.  However, they split
from each other approximately within $R/r_n\leq 1$ (with $r_n= n^2= 900$),
which can be seen from the sum of the probability densities of even-$l$ and
odd-$l$ states. They differ when the overlap
$\langle\psi^{B}_{n}|\psi^{A}_{n}\rangle$ is not exponentially small, see
Fig.~\ref{fig:sumpd}.  This feature is general for all principle quantum
numbers. The additional splitting also suggests that the system can be more
stable with the inclusion of more than one neutral perturber, a situation we
will investigate in more detail in Section \ref{sc:3gs}.

\begin{figure}
\includegraphics[width=1.0\columnwidth]{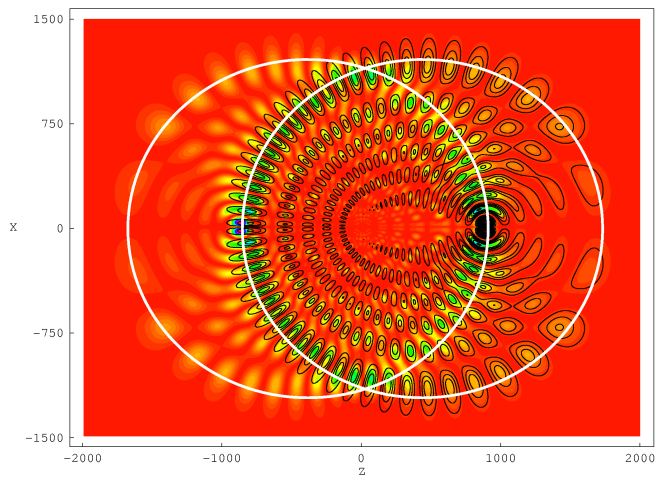}
\hspace*{0.05\textwidth}
\caption{\label{fig:2gspd}
Cut of the electronic probability density along the internuclear axis at the
deepest point of the potential well, $R=879$. The black contour lines show the
probibility density of the trilobite (diatomic) wavefunction, and the
background coloured plot is for the $N=2$ (collinear triatomic)
configuration. The ground-state atoms are located at $(x,z)=$$(0,\pm879)$ and
the Rydberg atom at $(0, 0)$. The two white solid lines show the classical
Kepler ellipses.}
\end{figure}

In Fig.~\ref{fig:2gspd}, we show the contour plot of the probability density
of the diatomic and triatomic systems at the interatomic distance $R_m=879$,
which corresponds to the deepest potential energy. The special minimum
configuration can be most clearly identified by means of the classical Kepler
orbits along which the two trilobite states of the molecule are scarred (see
also Ref.~\cite{Gra02}). Each Kepler ellipse has one ground-state atom in one
of its foci and touches the other ground-state atom.

\begin{figure}
\begin{center}
\mbox{
\subfigure[]{\label{fig:c3}\includegraphics[width=0.45\linewidth]{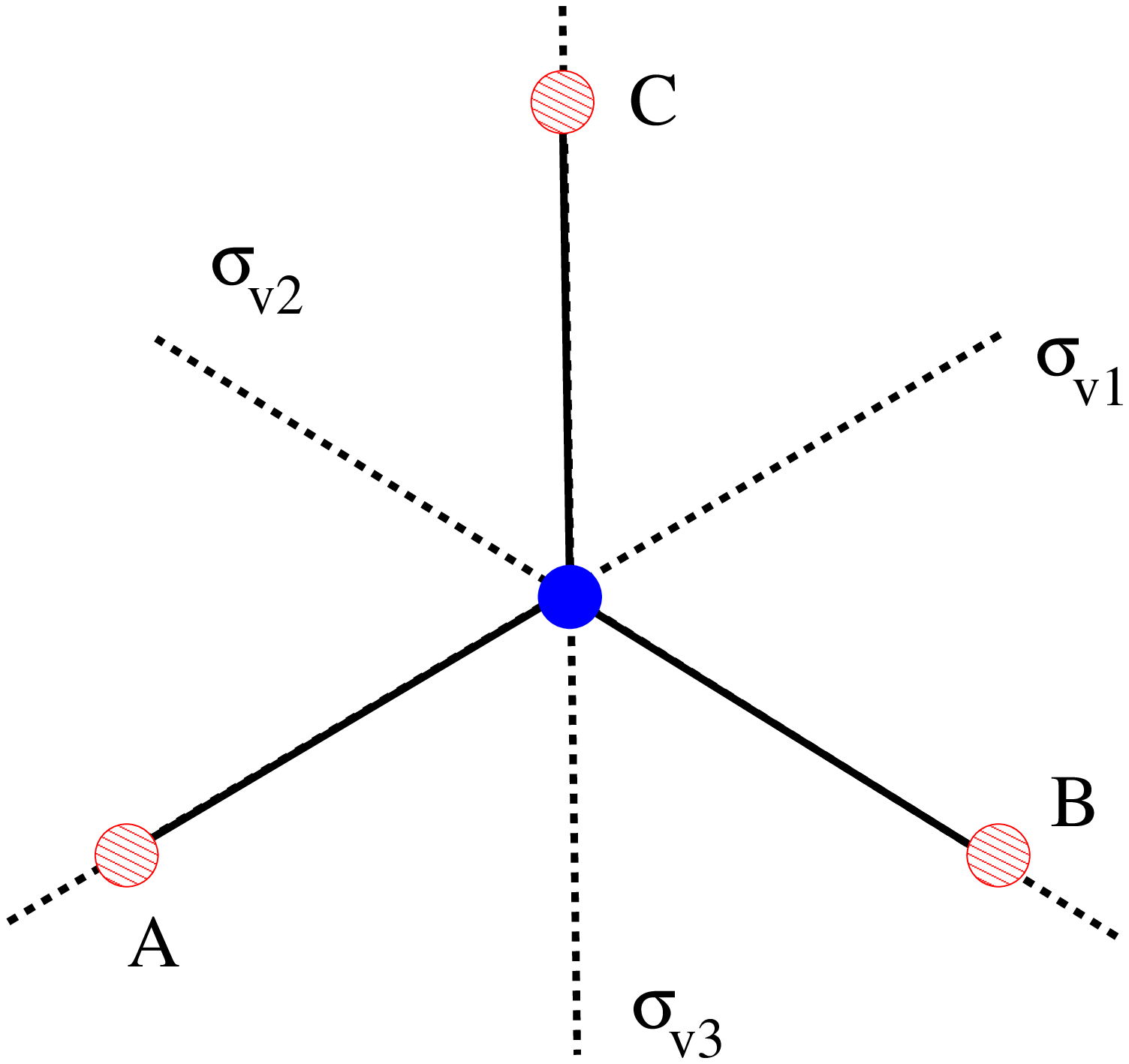}}
\subfigure[]{\label{fig:c4}\includegraphics[width=0.45\linewidth]{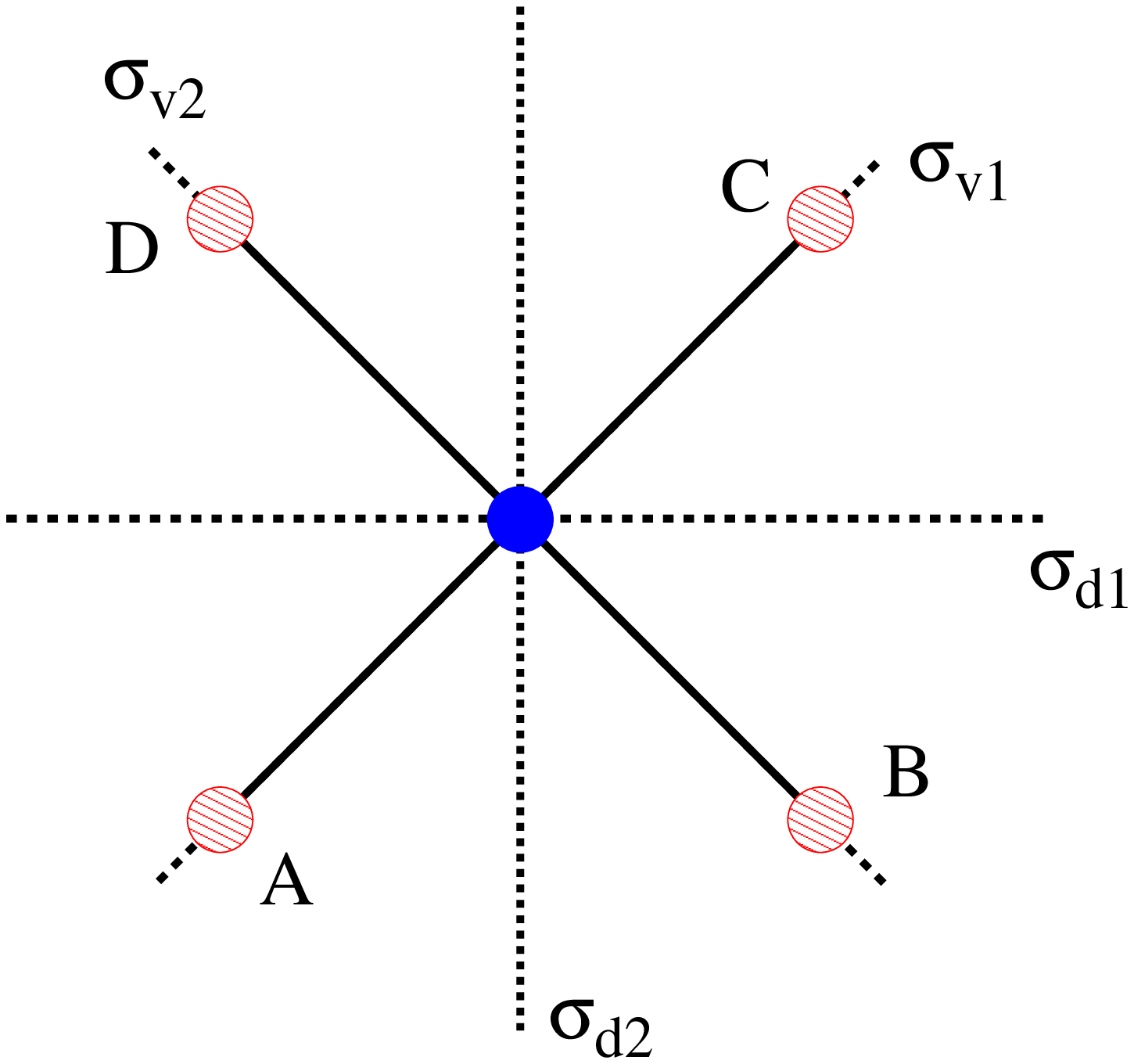}} 
}
\end{center}
\caption{\label{fig:geo}Illustration of the geometry of the (a) $C_{3,v}$ and (b)
$C_{4,v}$ configurations. The dashed lines are the planes of reflection. The shaded and the
solid circles are the ground-state and the Rydberg atoms respectively.}
\end{figure}

\section{\label{sc:3gs}Planar Polyatomic Molecules of Triangular and Quadratic Shape (N=3,4)}

\begin{figure}[b]
\includegraphics[width=1.0\columnwidth]{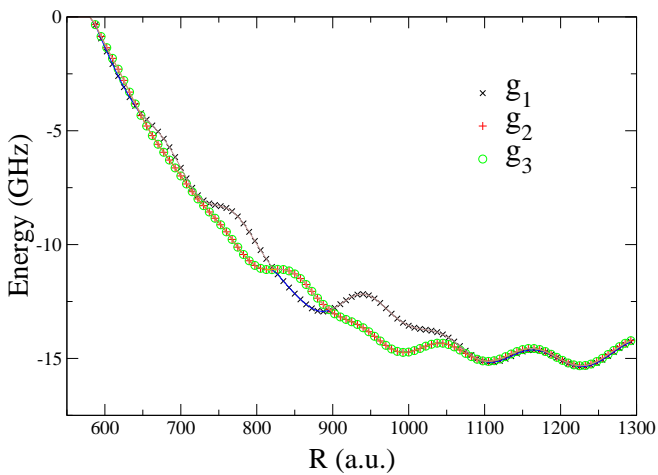}
\caption{\label{fig:3gsbo}
The three adiabatic energy curves $g_i$ for a triangular configuration as a
function of Rydberg--ground state atomic distance $R$ (see
Fig.~\ref{fig:geo}a). The coding of the data is analogous to
Fig.~\ref{fig:2gsbo}.}
\end{figure}

For $N\geq3$, we choose $\hat{z}$ to be perpendicular to the
plane containing the atoms. Figure \ref{fig:geo} illustrates the spacial geometry
of the complexes. The degenerate states that contribute, i.e., the
index $q$ in Eq.~\eqref{tri}, are the states with $l+m$ being an even integer,
exclusive of $l=0,1,2$ states. The results are shown in Fig.~\ref{fig:3gsbo}.

For the $N=3$ case, three energy eigenvalues split away from the $n$-manifold,
and as in the case of $N=2$, beyond $r_n$ they converge to the BO curve for
the dimer case. Note, however, that two of the energy levels are degenerate at
any given distance within $r_n$, indicating that there are additional
symmetries preserved under the perturbation of the ground-state atoms. To
elucidate these symmetries, we again use the trilobite state as basis
functions to construct the relevant symmetry-adapted orbitals.  But unlike in
the previous collinear configuration, where taking account of the parity as
the relevant symmetry is intuitive, we have to use a systematic approach for
$N$=3 or larger.  A method which has been used extensively to find the
symmetry-adapted orbitals is the \emph{projection operator method}, which is
expressed mathematically as \cite{Levine75}
\begin{equation}\label{pom}
g=\frac{l_{j}}{h}\sum_{\hat{R}}\chi_{j}^{*}(\hat{R})\hat{O}_{R}f\,,
\end{equation}
where $\hat{O}_R$ is the operator for a particular symmetry operation
$\hat{R}$, e.g., rotation or reflection, etc.. The $l_{j}$ and $\chi_j$ are,
respectively, the dimension and the character of the $j$-th irreducible
representation (irrep) of the symmetry group, to which the system belongs,
while $h$ is the order of the group.  The sum extends over all symmetry
operations in the group. This equation allows us to find the symmetry-adapted
functions $\{g_i\}$ from any original basis set $\{f_i\}$. The general proof
of this proceedure can be found, for example, in section 6.6 of
Ref.~\cite{Schon65}.

The name of this procedure originates from the fact that the pre-factor in
front of $f$ in the above equation can be viewed as a projection operator that
projects the basis set $\{f_{i}\}$ into a basis set $\{g_{i}\}$ that
diagonalizes the Hamiltonian matrix. In other words, $g$ and $f$ in \eq{pom}
are vectors, and the operator
\begin{equation}\label{eq:pj}
\hat{P}_{\Gamma}\equiv\frac{l_{j}}{h}\sum_{\hat{R}}\chi_{j}^{*}(\hat{R})\hat{O}_{R}
\end{equation}
can be represented by a unitary matrix.

The configurations of $N$=3 and 4 correspond to the symmetry groups $C_{3,v}$ and
$C_{4,v}$, and their character tables of the irreducible representations are shown in
Table~\ref{tb:ch}.

\begin{table}
\caption{\label{tb:ch}Character tables of the irreducible representations of (a) $C_{3,v}$ 
and (b) $C_{4,v}$ \cite{Salt72}. The labeling follows the conventional rules.}
\begin{center}
\mbox{
\subtable[]{\label{tb:c3}
\begin{tabular}{r|rrr}
	& E & $2C_3$ & $3\sigma_v$\\\hline
$\Gamma_{1}$ & 1 & 1& 1\\
$\Gamma_{2}$ & 1 & 1&-1\\
$\Gamma_{3}$ & 2 &-1& 0
\end{tabular}
}
\subtable[]{
\begin{tabular}{r|rrrrr}
	& E & $C_{2}$ & $2C_{4}$ & $2\sigma_{v}$ & $2\sigma_{d}$\\\hline
$\Gamma_{1}$ & 1 & 1 & 1 & 1 & 1\\
$\Gamma_{2}$ & 1 & 1 & 1 &-1 &-1\\
$\Gamma_{3}$ & 1 & 1 &-1 & 1 &-1\\
$\Gamma_{4}$ & 1 & 1 &-1 &-1 & 1\\
$\Gamma_{5}$ & 2 &-2 & 0 & 0 & 0
\end{tabular}\label{tb:c4}
}
}%mbox
\end{center}
\end{table}

We find that in the $N$=3 case, the representation of the symmetry operations using
the trilobite state as the basis set contains only two of the total three irreducible 
representations, $\Gamma_1$ and $\Gamma_3$ (see Table~\ref{tb:c3} and 
Appendix \ref{ap:n3} for details), which are one- and two-dimensional
respectively. The symmetry-adapted orbitals constructed will then, according to the
fundamental theory of quantum mechanics, consist of a non-degenerate and
two degenerate states. They are, repectively, $g_1$, $g_2$ and $g_3$ shown explicitly in
Eq.~\eqref{eq:N3g1}, \eqref{eq:N3g2} and \eqref{eq:N3g3} in Appendix \ref{ap:n3}.

The energy expectation values,
\begin{equation}
\langle \hat{V}_{N}(R)\rangle_{g_i}\equiv\langle g_i|\hat{V}_{N}(R)|g_i\rangle,
\end{equation}
calculated using the $g_i$-functions $i = 1,2,3$ are plotted in Fig.~\ref{fig:3gsbo}.
As expected, the two curves belonging to $\Gamma_3$ overlap with each other at 
all distances $R$.
Note that applying $\hat{P}_{\Gamma_2}$ onto any of the basis functions produces
zero, which is a general feature when the irrep is not contained in the overall
representation.

\begin{figure}
\includegraphics[width=1.0\columnwidth]{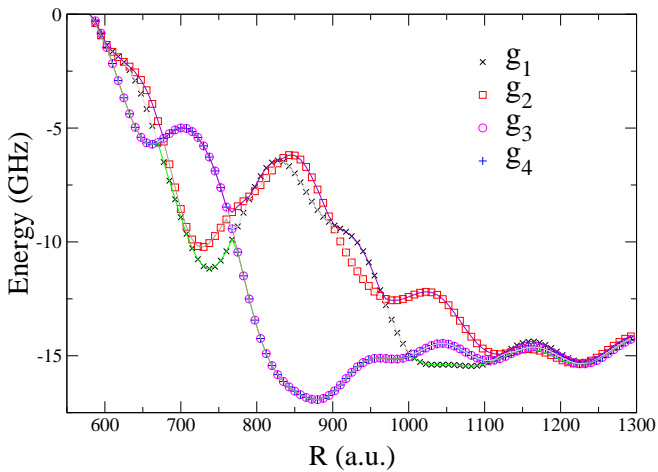}
\caption{\label{fig:4gsbo}Same as Fig.~\ref{fig:3gsbo}, but for a square geometry with a ground-state atom at each corner (see Fig.~\ref{fig:geo}b) with four energy curves $g_i$, see text.}
\end{figure}

The same analysis for the $C_{4,v}$ symmetry reveals that the overall
representation contains $\Gamma_1$, $\Gamma_3$, and $\Gamma_5$ (see
Table.~\ref{tb:c4}), with the first being one-dimensional, and the second and
the third two-dimensional (see Appendix
\ref{ap:n4} for detail). Hence,
there are two sets of doubly-degenerate BO curves and a non-degenerate
one. Applying Eq.~\eqref{pom}, we obtain the symmetry-adapted orbitals $g_i$,
$i=1,2,3,4$ as shown in Eq.~\eqref{wf:quad1}-\eqref{wf:quad4} in Appendix
\ref{ap:n4}. The adiabatic energy levels from the analytical and numerical
results are plotted in Fig.~\ref{fig:4gsbo}. Again, the graph shows a perfect
agreement between the two results.

\section{Conclusion}

We have used the Fermi pseudo-potential to model the effect neutral Rb
ground-state atoms have on a Rydberg electron. Cuts through the resulting
potential surface, adiabatic in the distance of the ground-state perturbers
from the ionic core of the Rydberg electron, have been calculated for
different arrangements of the ground-state atoms which form planar polyatomic
molecules. We found that more ground-state atoms lead to more deeply bound
mole\-cules compared to the original diatom as studied by Greene and coworkers
\cite{Greene00}. A systematic understanding of the structure and symmetry of
such molecules can be gained by taking the trilobite (diatomic) wavefunction
as a basic unit and constructing symmetry-adapted orbitals as demonstrated in
Section
\ref{sc:2gs} and \ref{sc:3gs}. For two ground-state atoms the procedure is relative simple
and intuitive, while three or more ground-state atoms require a systematic
approach, such as the projection operator method.

In the case where $N$ is larger than the number of degenerate states $q_\mathrm{max}$, our
method of constructing the perturbed wavefunction should still work, and will yield
only $q_\mathrm{max}$ linearly independent states.

The effect of $p$-wave electron scattering plays an important role especially
in hydrogen-like atoms. Our calculations can be extended into the case of higher
partial wave scattering by using the appropriate pseudo-potentials
formulated by Omont \cite{Omont77}. Previous studies \cite{Ham02,Khu02} for
dimers show that the potential curve of the $p$-wave scattering crosses the
potential well at $R\approx 1200$, and could potentially destabilize the
trilobite while also providing additional potential wells. It will be also
interesting to see how the spatial arrangement of the atoms affect the energy
of the system in this process, since the potential is now dependent on the
gradient of the electronic wavefunction in the 3D space.

The present work is a first exploration of the possibility to form polyatomic
molecules from a single Rydberg atom and a number of ground-state atoms. We
have only determined a cut (at equal distances of the ground-state atoms to
the Rydberg atom) through the multidimensional potential surface which
resembles the potential for the breathing mode of the molecule. Future
analysis and realistic assessment of the quantitative features of such species
must include the vibrational motion of the atoms.

\appendix

\section{Derivation of the Symmetry-adapted Orbitals}

\subsection{\label{ap:n3}Planar Polyatomic Molecule with N=3}

The molecule formed in this configuration has the
symmetry of the point group $C_{3,v}$. Using trilobite states
as the basis set to contruct the corresponding representation,
one obtains the following matrices,
\begin{subequations}\label{eq:c3}
\begin{align}
	\hat{O}_{E} & =
	\begin{pmatrix}
	1 & 0 & 0\\
	0 & 1 & 0\\
	0 & 0 & 1
	\end{pmatrix}&
		\hat{O}_{{C}_{3}^1} & =
		\begin{pmatrix}
		0 & 1 & 0\\
		0 & 0 & 1\\
		1 & 0 & 0
		\end{pmatrix}\\
	\hat{O}_{{C}_{3}^2} & =
	\begin{pmatrix}
	0 & 0 & 1\\
	1 & 0 & 0\\
	0 & 1 & 0
	\end{pmatrix}&
		\hat{O}_{{\sigma}_{v1}} & =
		\begin{pmatrix}
		1 & 0 & 0\\
		0 & 0 & 1\\
		0 & 1 & 0
		\end{pmatrix}\\
	\hat{O}_{{\sigma}_{v2}} & =
	\begin{pmatrix}
	0 & 0 & 1\\
	0 & 1 & 0\\
	1 & 0 & 0
	\end{pmatrix}&
		\hat{O}_{{\sigma}_{v3}} & =
		\begin{pmatrix}
		0 & 1 & 0\\
		1 & 0 & 0\\
		0 & 0 & 1
		\end{pmatrix}
\end{align}
\end{subequations}

In the above notations, $E$ is the identity; $C_{k}^p$ denotes the rotation about
$\hat{z}$-axis by angle $2\pi p/k$;
and the $\sigma$'s are
the relfections through the planes perpendicular to the plane of the atoms,
as indicated in Fig.~\ref{fig:c3}.

It is clear that when one of the above operators, say $\hat{O}_{{C}_{3}^1}$, is 
applied on the original vector, the result is a $120^{\circ}$ rotation
about $\hat{z}$ in the counter-clockwise direction, assuming that $\hat{z}$
is pointing perpendicularly into the paper, i.e.
\begin{equation}
\begin{pmatrix}
0 & 1 & 0\\
0 & 0 & 1\\
1 & 0 & 0
\end{pmatrix}
\begin{pmatrix}
\psi_{n}^{A}\\
\psi_{n}^{B}\\
\psi_{n}^{C}
\end{pmatrix}\longrightarrow
\begin{pmatrix}
\psi_{n}^{B}\\
\psi_{n}^{C}\\
\psi_{n}^{A}
\end{pmatrix}.
\end{equation}

From Eq.~\eqref{eq:c3}, the character of the representation in such basis set can
be determined by taking the trace of each matrix, and they are summarized in the
table below:
\begin{center}
\begin{tabular}{c|ccc}
& E & $2C_{3}$ & $3\sigma_v$ \\\hline
$\Gamma_\mathrm{total}$ & 3 & 0  & 1
\end{tabular}
\end{center}
Here, we have used $\Gamma_\mathrm{total}$ to denote the representation formed by the trilobite
states. By inspecting the character table of the irreps of $C_{3,v}$
(Table~\ref{tb:c3}), one immediately sees that the current representation
is a direct sum of $\Gamma_1$ and $\Gamma_3$, namely,
\begin{equation}
\Gamma_\mathrm{total} = \Gamma_1 \oplus \Gamma_3.
\end{equation}
Now, we determine the projection operators in each irrep by using Eq.~\eqref{eq:pj}.
The order of the group is $h=6$, and the dimensions for $\Gamma_1$ and $\Gamma_3$
are $l_1=1$ and $l_3=2$, respectively. Equation \eqref{eq:pj} then yields
\begin{subequations}
\begin{align}
\hat{P}_{\Gamma_1} & = \frac{1}{6}\left[\hat{O}_E + \hat{O}_{C_{3}^1} +
\hat{O}_{C_{3}^2} + \hat{O}_{\sigma_{v1}} + \hat{O}_{\sigma_{v2}} + \hat{O}_{\sigma_{v3}}
\right] \label{eq:pj3a},\\
\hat{P}_{\Gamma_2} & = \frac{1}{6}\left[\hat{O}_E + \hat{O}_{C_{3}^1} +
\hat{O}_{C_{3}^2} - \hat{O}_{\sigma_{v1}} - \hat{O}_{\sigma_{v2}} - \hat{O}_{\sigma_{v3}}
\right],\\
\hat{P}_{\Gamma_3} & = \frac{1}{6}\left[2\hat{O}_E + \hat{O}_{C_{3}^1} +
\hat{O}_{C_{3}^2}\right]\label{eq:pj3c}.
\end{align}
\end{subequations}

Since $\Gamma_\mathrm{total}$ contains only $\Gamma_1$ and $\Gamma_3$, we need to apply
only Eq.~\eqref{eq:pj3a} and \eqref{eq:pj3c} to our basis set in order to obtain the
symmetry-adapted orbitals. Acting the trivial operator $\hat{P}_{\Gamma_1}$ on
the trilobite wavefunction $|\psi_{n}^{A}\rangle$, we obtain the first symmetry-adapted
orbital
\begin{equation}\label{eq:N3g1}
\hat{P}_{\Gamma_1}f_A = \hat{P}_{\Gamma_1} |\psi_{n}^{A}\rangle 
= \frac{1}{3}\left[ |\psi_{n}^{A}\rangle
+ |\psi_{n}^{B}\rangle + |\psi_{n}^{C}\rangle \right] \equiv g_1.
\end{equation}
The same equations are obtained if one acts $\hat{P}_{\Gamma_1}$ on 
$|\psi_{n}^{B}\rangle$ or $|\psi_{n}^{C}\rangle$ which are obviously
linearly-dependent. However, acting $\hat{P}_{\Gamma_3}$
on $|\psi_{n}^{A}\rangle$ , $|\psi_{n}^{B}\rangle$  and $|\psi_{n}^{C}\rangle$
gives, repectively,
\begin{subequations}
\begin{align}\label{eq:N3g2}
\hat{P}_{\Gamma_3} |\psi_{n}^{A}\rangle & = \frac{1}{6}\left[ 2|\psi_{n}^{A}\rangle
- |\psi_{n}^{B}\rangle - |\psi_{n}^{C}\rangle \right] \equiv g_2,\\
\hat{P}_{\Gamma_3} |\psi_{n}^{B}\rangle & = \frac{1}{6}\left[ 2|\psi_{n}^{B}\rangle
- |\psi_{n}^{C}\rangle - |\psi_{n}^{A}\rangle \right] \label{eq:sym3},\\
\hat{P}_{\Gamma_3} |\psi_{n}^{C}\rangle & = \frac{1}{6}\left[ 2|\psi_{n}^{C}\rangle
- |\psi_{n}^{B}\rangle - |\psi_{n}^{A}\rangle \right]\label{eq:sym4}.
\end{align}
\end{subequations}

Since $\Gamma_3$ is a three-dimensional irrep, two of the above equations can be
combined, by subtracting Eq.~\eqref{eq:sym3} by \eqref{eq:sym4}, giving
\begin{equation}
\frac{1}{6}\left[|\psi^{B}_{n}(\vec{r})\rangle-|\psi^{C}_{n}(\vec{r})
\rangle\right]\label{eq:N3g3} \equiv g_3,
\end{equation}
so that finally we have three linearly-independent wave functions, which we 
call $g_1$, $g_2$ and $g_3$.

\subsection{\label{ap:n4}Planar Polyatomic Molecule with N=4}

Following the same procedure as in the case of $N=3$, one finds the matrices
of the symmetry operations in the point group $C_{4,v}$ in the present basis set as,
\begin{subequations}\label{eq:c4}
\begin{align}
	\hat{O}_{E} & =
	\begin{pmatrix}
	1 & 0 & 0 & 0\\
	0 & 1 & 0 & 0\\
	0 & 0 & 1 & 0\\
	0 & 0 & 0 & 1
	\end{pmatrix}&
		\hat{O}_{{C}_{2}^1} & =
		\begin{pmatrix}
		0 & 0 & 1 & 0\\
		0 & 0 & 0 & 1\\
		1 & 0 & 0 & 0\\
		0 & 1 & 0 & 0
		\end{pmatrix}\\
	\hat{O}_{{C}_{4}^1} & =
	\begin{pmatrix}
	0 & 1 & 0 & 0\\
	0 & 0 & 1 & 0\\
	0 & 0 & 0 & 1\\
	1 & 0 & 0 & 0
	\end{pmatrix}&
		\hat{O}_{{C}_{4}^3}& =
		\begin{pmatrix}
		0 & 0 & 0 & 1\\
		1 & 0 & 0 & 0\\
		0 & 1 & 0 & 0\\
		0 & 0 & 1 & 0
		\end{pmatrix}\\
	\hat{O}_{{\sigma}_{v1}} & =
	\begin{pmatrix}
	1 & 0 & 0 & 0\\
	0 & 0 & 0 & 1\\
	0 & 0 & 1 & 0\\
	0 & 1 & 0 & 0
	\end{pmatrix}&
		\hat{O}_{{\sigma}_{v2}} & =
		\begin{pmatrix}
		0 & 0 & 1 & 0\\
		0 & 1 & 0 & 0\\
		1 & 0 & 0 & 0\\
		0 & 0 & 0 & 1
		\end{pmatrix}\\
	\hat{O}_{{\sigma}_{d1}} & =
	\begin{pmatrix}
	0 & 1 & 0 & 0\\
	1 & 0 & 0 & 0\\
	0 & 0 & 0 & 1\\
	0 & 0 & 1 & 0
	\end{pmatrix}&
		\hat{O}_{{\sigma}_{d2}} & =
		\begin{pmatrix}
		0 & 0 & 0 & 1\\
		0 & 0 & 1 & 0\\
		0 & 1 & 0 & 0\\
		1 & 0 & 0 & 0
		\end{pmatrix}
\end{align}
\end{subequations}
where the notations are as before, and the planes of reflections are indicated
in Fig.~\ref{fig:c4}.

The character of $\Gamma_\mathrm{total}$ can again be determined by taking the trace of
each matrix above, and they are:
\begin{center}
\begin{tabular}{c|ccccc}
& E & $ C_{2}$ & $2C_4$ & $2\sigma_v$ & $2\sigma_d$ \\\hline
$\Gamma_\mathrm{total}$ & 4 & 0  & 0 & 2 & 0
\end{tabular}
\end{center}

Again, from the character table of the irrep (Table \ref{tb:c4}), one finds 
that the  representation $\Gamma_\mathrm{total}$ is a direct sum of
\begin{equation}
\Gamma_\mathrm{total} = \Gamma_1 \oplus \Gamma_3 \oplus \Gamma_5.
\end{equation}
Hence, we know that in this representation there are two one-dimensional
and one two-dimensional irreps. Their corresponding projection operators can
be obtained by applying Eq.~\eqref{eq:pj}, where in this case, $h=8$, and $l_1$,
$l_3$ and $l_5$ are 1, 1 and 2, repectively. Therefore, the projection operators
are
\begin{subequations}
\begin{align}
\begin{split}
	\hat{P}_{\Gamma_1} & = \frac{1}{8}[\hat{O}_E + \hat{O}_{C_{2}^1} +
							\hat{O}_{C_{4}^1} + \hat{O}_{C_{4}^2}\\
					& \quad + \hat{O}_{\sigma_{v1}} + \hat{O}_{\sigma_{v2}}
						+ \hat{O}_{\sigma_{d1}} + \hat{O}_{\sigma_{d2}}],
\end{split}\\
\begin{split}
	\hat{P}_{\Gamma_3} & = \frac{1}{8}[\hat{O}_E + \hat{O}_{C_{2}^1} -
							\hat{O}_{C_{4}^1} - \hat{O}_{C_{4}^2} \\
					& \quad + \hat{O}_{\sigma_{v1}} + \hat{O}_{\sigma_{v2}}
						- \hat{O}_{\sigma_{d1}} - \hat{O}_{\sigma_{d2}}],
\end{split}\\
\hat{P}_{\Gamma_5} & = \frac{1}{4}\left[\hat{O}_E - \hat{O}_{C_{2}^1}\right].
\end{align}
\end{subequations}

The symmetry-adapted orbitals can then be obtained in a similar way as in Appendix \ref{ap:n3},
which yields the following four linearly-independent equations:
\begin{align}
g_1 &= \frac{1}{4}\left[|\psi^{A}_{n}(\vec{r})\rangle+|\psi^{B}_{n}(\vec{r})\rangle
+|\psi^{C}_{n}(\vec{r})\rangle+|\psi^{D}_{n}(\vec{r})\rangle\right]\label{wf:quad1},\\
g_2 &= \frac{1}{4}\left[|\psi^{A}_{n}(\vec{r})\rangle-|\psi^{B}_{n}(\vec{r})\rangle
+|\psi^{C}_{n}(\vec{r})\rangle-|\psi^{D}_{n}(\vec{r})\rangle\right]\label{wf:quad2},\\
g_3 &= \frac{1}{4}\left[|\psi^{A}_{n}(\vec{r})\rangle-|\psi^{C}_{n}(\vec{r})\rangle
\right]\label{wf:quad3},\\
g_4 &= \frac{1}{4}\left[|\psi^{B}_{n}(\vec{r})\rangle-|\psi^{D}_{n}(\vec{r})\rangle
\right]\label{wf:quad4}.
\end{align}

\bibliographystyle{epj}
\bibliography{literature}
\end{document}